\documentclass[aip,pop,reprint,10pt,superscriptaddress,showpacs]{revtex4-1}
\usepackage{amsfonts} 
\usepackage{amsmath}
\usepackage{amssymb}
\usepackage{graphicx}
\usepackage{subfigure}
\usepackage{color}
\newcommand*\xbar[1]{%
  \hbox{%
    \vbox{%
      \hrule height 0.5pt 
      \kern0.5ex
      \hbox{%
        \kern-0.1em
        \ensuremath{#1}%
        \kern-0.1em
      }%
    }%
  }%
} 
\begin{document}
\title{Elliptically polarized electromagnetic waves in a magnetized quantum electron-positron plasma with effects of  exchange-correlation}
\author{M. Shahmansouri}
\email{mshmansouri@gmail.com}
\affiliation{Department of Physics, Faculty of Science, Arak University, Arak 38156- 8 8349, Iran}
\author{A. P. Misra}
\email{apmisra@visva-bharati.ac.in; apmisra@gmail.com}
\affiliation{Department of Mathematics, Siksha Bhavana, Visva-Bharati University, Santiniketan-731 235, West Bengal, India}
\pacs{}

\begin{abstract}
The dispersion properties of  elliptically polarized electromagnetic (EM) waves in a magnetized  electron-positron-pair (EP-pair) plasma   are studied with the effects of particle dispersion associated with the Bohm potential, the Fermi degenerate pressure, and the exchange-correlation force.  Two possible  modes of the extraordinary or X wave, modified by these  quantum effects, are identified and their propagation characteristics  are investigated numerically. It is shown that  the upper-hybrid frequency,  and the cutoff and resonance frequencies are no longer constants but are dispersive due to these quantum effects.   It is found that the particle dispersion and the exchange-correlation force can have different dominating roles on each other depending on  whether the  X waves  are of short or  long wavelengths (in comparison with the Fermi Debye length).  The present investigation should be useful for understanding the collective behaviors of EP plasma oscillations and the propagation of  extraordinary waves in magnetized  dense EP-pair plasmas. 
\end{abstract}
\date{11 April 2016}
\maketitle
\section{Introduction} \label{sec-introduct}
Study of electron-positron (EP) plasmas provides  important insights about the early universe \cite{ress1983,misner1973} and astrophysical objects  (such as supernova remnants, pulsars,   active galactic nuclei etc.) \cite{miller1987,goldreich1969,michel1982}. Such plasmas  may be created by different physical mechanisms, for example, collisions between the accelerated particles, streaming of charged particles along the curved magnetic fields and the interaction of high-energy lasers with plasmas. 
Over the past few years, a number of experiments have  been proposed to create EP plasmas (see, e.g., Refs. \onlinecite{surko1989,greaves1995,boehmer1995}). However,  because of the fast annihilation and the formation of positronium atoms, the identification of collective modes in these EP plasmas may be practically impossible. To resolve this issue, some attempts have been made to produce    high-density $(\sim10^{16}$ cm$^{-3})$  neutral EP plasmas (ion-free plasmas with unique characteristics)   in the laboratory \cite{sarri2015}.
\par
  On the other hand, it has been shown that the propagation of electromagnetic (EM) radiation in EP plasmas, is responsible for the high effective temperatures of pulsar radio emissions \cite{lominadze1982,lominadze1983,gedalin1985}.  The knowledge on the dynamics of such  EM waves in EP plasmas is essential for understanding the radiation properties of astrophysical objects, even the media exposed to the field  of super-strong laser radiation \cite{tatsuno2003}. These may be the reason behind the attraction of attention on the investigation of collective phenomena in EP plasmas.  
Several authors  have investigated the propagation characteristics of EM waves in EP plasmas (see, e.g., Refs. \onlinecite{yu1984,stenflo1985,yu1986,marklund2006,liu2010}). It is well known that the quantum effects can play a vital role in plasmas when the de Broglie wavelength becomes comparable to the average inter-particle distance \cite{manfredi2005}. In such  systems, the quantum effects significantly change the collective behaviors of plasma species \cite{gardner1996,manfredi2001,shukla2006,misra2010a}. For instance, the Bohm potential leads to appearance of higher-order corrections in the dispersion relation, meanwhile the spin magnetization contributes to the linear dispersion of electromagnetic modes \cite{shukla2011,misra2010b} etc. Several theoretical models have been proposed to study the collective behaviors and associated nonlinear structures in quantum plasmas \cite{shukla2011}. Furthermore, the quantum magnetohydrodynamic (QMHD) model   has also been extensively used to study the influence of quantum effects on the wave propagation in quantum plasmas \cite{misra2010b,stefan2010,mushtaq2012,shukla2009,masood2010}.   Shukla \cite{shukla2006}  investigated the combined effects of the quantum Bohm potential and the electron spin-$1/2$  effects on the extraordinary (X) EM (X-EM) waves in a warm dense magnetoplasma. He showed that the quantum effects significantly modify the dispersion properties of such waves. The influence of electron magnetization spin current on the X-EM waves in a magnetized quantum plasma has also  been examined by Li {\it et. al.} \cite{li2012}. 
\par
 It has been shown that the  effects of exchange and correlation forces  on the plasma collective oscillations can be significant and can even play dominant roles over the particle dispersion \cite{zamanian2015}.
To complement and give new insights into the previously published works \cite{li2012}, we propose here to address the propagation properties of X-EM waves in a magnetized quantum EP plasma including the combined influences of the exchange-correlation potential and the Bohm potential. In order to describe the present plasma system, we employ the QMHD model, and obtain the modified dispersion relation. It is shown that the quantum effects modify the dispersion properties  of X-EM waves together with   the cutoff and resonance frequencies. The manuscript is organized as follows: The basic equations governing dynamics of EP-pair plasmas   are given in Sec. \ref{sec-th-model}. The dispersion relation   of the X-EM waves is derived   in Sec. \ref{sec-disper-relation}. Our theoretical results and a brief discussion are provided in  Sec. \ref{sec-results-discuss}. Finally, we conclude our results in Sec. \ref{sec-conclusion}.
\section{Theoretical model}\label{sec-th-model}
We consider a magnetized quantum plasma consisting  of degenerate electrons as well as positron fluids  in   presence of an uniform external  magnetic field acting along  the $z$-axis, i.e., ${\bf B}_0=B_0\hat{z}$. The QMHD model (which can be obtained from the self-consistent Hartree equations \cite{manfredi2005} or from the phase-space Wigner-Poisson equations \cite{manfredi2007})  describing the dynamical behaviors of electrons and positrons  is given by
\begin{equation}
\frac{\partial n_j}{\partial t}+\nabla\cdot(n_j {\bf v}_j)=0, \label{cont-eq}
\end{equation}
\begin{equation}
m\frac{\partial {\bf v}_j}{\partial t}={q_j}\left({\bf E}+{\bf v}_j\times {\bf B}\right)-\frac{\nabla P_j}{n_j}-\nabla V_{qj}-\nabla V_{xj},\label{moment-eq}
\end{equation}
where $n_j$, ${\bf v}_j$ and $q_j$, respectively, denote the number density, velocity and charge of $j$-species particle, and $m_e=m_p=m$ is the mass of electrons $(j=e)$ and positrons $(j=p)$. Also,   $ {\bf E}~({\bf B})$ is the electric (magnetic) field and $\epsilon_0$ is the permittivity of free space. The equations  \eqref{cont-eq}-\eqref{moment-eq} are   closed by the Maxwell's equations, to be presented shortly. Furthermore,  $P_j$ is the weakly relativistic  pressure for degenerate electrons and positrons given by \cite{chandrasekhar1935}
\begin{equation}
P_j=\frac{1}{5}\frac{m V_F^2}{n_0^{2/3}}n_j^{5/3}, \label{pressure-ep}
\end{equation}
where   $V_{F}\equiv\sqrt{2k_BT_{F}/m}=(\hbar/m)(3\pi^2n_{0})^{1/3}$ is the Fermi velocity with $k_B$ denoting the Boltzmann constant, $T_F$ the Fermi temperature and $n_0$ the equilibrium number density of electrons and positrons. The particle dispersion associated with the density correlation due to quantum fluctuation (tunneling) is included in Eq. \eqref{moment-eq} via the Bohm potential $V_{qj}$ given by \cite{manfredi2005}
\begin{equation}
V_{qj}=-\frac{\hbar^2}{2m}\frac{1}{\sqrt{n_j}}\nabla^2\sqrt{n_j}. \label{bohm-potential}
\end{equation}  
 Furthermore, in dense plasmas the electron/positron exchange/correlation effects play a non-negligible role \cite{zamanian2015}. The general framework that produces such exchange-correlation potentials is the density functional theory (DFT) \cite{dreizler1990}. This potential is a complicated function of the particle number density given by \cite{hedin1971,brey1990,crouseilles2008,shahmansouri2015}
\begin{eqnarray}
V_{xj}=&&-0.985\left( n_j^{1/3}e^2/\epsilon\right)\left[1+\left(0.034/n_j^{1/3}a_B\right)\right.\notag\\
&&\left. \times \ln\left(1+18.376n_j^{1/3}a_B\right)\right], \label{exchange-corr}
\end{eqnarray}
where $\epsilon$ is the effective dielectric constant of the medium and $a_B=\epsilon\hbar^2/me^2$ is the Bohr radius. 
\par
 Next, linearizing the set of equations \eqref{cont-eq} and \eqref{moment-eq}  by splitting up the physical quantities into their equilibrium (with a zero value or a quantity with suffix $0$) and perturbation parts (with suffix $1$), i.e.,  $n\sim n_0+n_1$, ${\bf v}\sim{\bf0}+{\bf v}_1$, ${\bf E}\sim{\bf 0}+{\bf E}_1$ and ${\bf B}\sim {\bf B}_0+{\bf B}_1$, we obtain (after omitting the suffix $1$)
 \begin{equation}
\frac{\partial n_j}{\partial t}+n_0 \nabla\cdot{\bf v}_j=0, \label{cont-eq-linear}
\end{equation}
\begin{equation}
\frac{\partial {\bf v}_j}{\partial t}=\frac{q_j}{m}\left({\bf E}+{\bf v}_j\times {\bf B}_0\right)- \frac{1}{3}\frac{\alpha}{n_0}\nabla n_j+\frac{\beta}{n_0}\nabla\left(\nabla^2n_j\right),\label{moment-eq-linear}
\end{equation}
where   $\beta=\hbar^2/4m^2$ is the coefficient of the term associated with the Bohm potential and
\begin{equation}
\alpha=V_{F}^2\left[3-(3\pi^2)^{2/3}H^2\left(0.985+\frac{0.616}{1+1.9/H^2}\right)\right], \label{alpha-ex-corr}
\end{equation}
 in which the first term in the square brackets appears due to the weakly relativistic degenerate pressure of electrons and positrons, and   the second term $\propto H$ is  due to the exchange and correlation effects. Here,  $H=\hbar\omega_{p}/mV_{F}^2\equiv\hbar\omega_{p}/2k_BT_{F}$ is the dimensionless  quantum parameter measuring the ratio of the electron/positron plasmon energy to the Fermi energy densities.  Note that for high-density plasmas,   $H\lesssim1$, and so the expression \eqref{alpha-ex-corr}  can be further simplified as
\begin{equation}
\alpha\approx V_{F}^2\left[3-(3\pi^2)^{2/3}H^2\right]. \label{alpha-ex-corr-approx}
\end{equation}
The perturbed electromagnetic fields are governed by the Maxwell's equations
\begin{equation}
\nabla\times{\bf E}=-\frac{\partial {\bf B}}{\partial t}, \label{faraday-law}
\end{equation}
\begin{equation}
\nabla\times{\bf B}=\mu_0{\bf J}+\frac{1}{c^2}\frac{\partial {\bf E}}{\partial t}, \label{ampere-law}
\end{equation}
where ${\bf J}=en_0({\bf v}_p-{\bf v}_e)$ is the current density. In what follows, we consider an elliptically polarized electromagnetic wave propagating along the $x$-axis. We suppose  that the polarized electric field lies in $xy$-plane, i.e.,  ${\bf E}=E_x\hat{x}+E_y\hat{y}$, and as before the external magnetic field is along the $z$-axis, i.e., ${\bf B}_0=B_0\hat{z}$. Assuming the  variation of the perturbed quantities to be of the form $\sim\exp(ikx-i\omega t)$ with $k~(\omega)$ denoting the wave number (frequency) of perturbations, we obtain from Eqs. \eqref{cont-eq-linear} and \eqref{moment-eq-linear} the following equations
\begin{equation}
\omega n_j=n_0{\bf k}\cdot{\bf v}_j, \label{nj}
\end{equation}
\begin{equation}
\omega{\bf v}_j=i\frac{q_j}{m}\left({\bf E}+{\bf v}_j\times{\bf B}_0\right)+\frac{\zeta k}{n_0}n_j\hat{x}, \label{vj}
\end{equation}
where $\zeta=V_{F}^2\left(1-3.2H^2+\frac{1}{4}H^2k^2\lambda_{F}^2\right)$ with $\lambda_F=V_F/\omega_p$ denoting the Fermi Debye length. 
Using  Eq. \eqref{nj} and separating the $x$- and $y$-components of Eq. \eqref{vj} we obtain
\begin{equation}
\left(1-\zeta\frac{ k^2}{\omega^2}\right)v_{jx}\pm\frac{i\omega_c}{\omega}v_{jy}=\mp\frac{ie}{m\omega}E_x, \label{vx}
\end{equation}
\begin{equation}
\frac{i\omega_c}{\omega}v_{jx}\mp v_{jy}=\frac{ie}{m\omega}E_y, \label{vy}
\end{equation}
where the upper and lower signs in $\pm$ or $\mp$ refer  to the signs of charges of electrons and positrons respectively. Next, from Eqs. \eqref{faraday-law} and \eqref{ampere-law} we obtain
\begin{equation}
i\omega E_x=\frac{m}{e}\omega_p^2\left(v_{px}-v_{ex}\right), \label{faraday-reduced}
\end{equation}
\begin{equation}
i\left(\omega^2-c^2k^2\right)E_y=\frac{m}{e}\omega_p^2\left(v_{py}-v_{ey}\right).\label{ampere-reduced}
\end{equation}
From Eqs. \eqref{vx} and \eqref{vy} solving for $v_{jx}$ and $v_{jy}$ we obtain
\begin{equation}
v_{jx}=\frac{(e/m)\omega}{\omega^2-\omega_c^2-\zeta k^2}\left(\pm iE_x-\frac{\omega_c}{\omega}E_y\right), \label{vjx}
\end{equation}
\begin{equation}
v_{jy}=\frac{(e/m)}{\omega^2-\omega_c^2-\zeta k^2}\left(\omega_cE_x\pm i\frac{\omega^2-\zeta k^2}{\omega}E_y\right). \label{vjy}
\end{equation}
\section{Dispersion relation} \label{sec-disper-relation}  Substituting the expressions of $v_{jx}$ and $v_{jy}$ from Eqs. \eqref{vjx} and \eqref{vjy} into Eqs. \eqref{faraday-reduced} and \eqref{ampere-reduced}, we obtain for nonzero values of $E_x$ and $E_y$ the following dispersion relations for the quantum modified X-wave and the upper-hybrid wave
\begin{equation}
\frac{c^2k^2}{\omega^2}=1-\frac{2\omega_p^2}{\omega^2}\left(\frac{\omega^2-\zeta k^2}{\omega^2-\omega_{c}^2-\zeta k^2}\right), \label{dispersion-x-wave}
\end{equation}
\begin{equation}
\omega^2=\omega_{uh}^2+\zeta k^2, \label{upper-hybrid-dispersion}
\end{equation}
where $\omega_{uh}=\sqrt{2\omega_p^2+\omega_c^2}$ is the upper-hybrid oscillation frequency in classical plasmas. 
 By disregarding the quantum effects $\propto \zeta$, one can obtain from Eq. \eqref{dispersion-x-wave} the  dispersion relation for   X-EM waves in electron-positron plasmas as
\begin{equation}
\frac{c^2k^2}{\omega^2}=1-\frac{2\omega_p^2}{\omega^2}\left(\frac{\omega^2}{\omega^2-\omega_{c}^2}\right),. \label{disp-X-classi-EP}  
\end{equation}
and from Eq. \eqref{upper-hybrid-dispersion} the usual hybrid frequency $\omega=\omega_{uh}$.
Also, by neglecting the quantum effects and the positron dynamics, and assuming positive ions to form the background plasma, one can obtain the known classical results for the X-wave as \cite{goldston1995}
  \begin{equation}
\frac{c^2k^2}{\omega^2}=1-\frac{\omega_p^2}{\omega^2}\left(\frac{\omega^2-\omega_p^2}{\omega^2-\omega_{uh}^2}\right). \label{disp-X-classi-E}  
\end{equation}
Now, comparing Eq. \eqref{dispersion-x-wave} with  Eq. \eqref{disp-X-classi-E}, we find that the dispersion of the X-EM wave in  EP plasmas is greatly  modified by the quantum effects $\propto \zeta$. Here, in Eq. \eqref{dispersion-x-wave}, the factor $2$ appears due to pair plasmas  with equal mass and unperturbed number density.   Also, after substitutions of the solutions  \eqref{vjx} and \eqref{vjy} into Eqs. \eqref{faraday-reduced} and \eqref{ampere-reduced}, we note that (because of  pair plasmas) the corresponding terms $\propto E_y$ and $E_x$ in these equations cancel each other. As a result the  term $\omega_p^2$ disappears from  both  the numerator and denominator of the term in the parentheses in Eq. \eqref{dispersion-x-wave}.   
  Furthermore, the dispersion equation \eqref{dispersion-x-wave}, upon dropping the positron dynamics (hence the corresponding factor  and the terms as above) and the contribution from the exchange correlation force, agrees with that in Ref. \onlinecite{li2012} except the spin magnetization current which has not been considered in the present model. 
\par
In what follows, we recast Eq. \eqref{dispersion-x-wave}   into its dimensionless form as
\begin{equation}
\frac{c_0^2K^2}{\Omega^2}=1-\frac{2}{\Omega^2}\frac{\Omega^2-\zeta_0 K^2}{\Omega^2-\Omega_c^2-\zeta_0 K^2}, \label{dispersion-x-wave-dimless}
\end{equation}
where $c_0=c/V_F,~\zeta_0=\zeta/V_F^2,~K=k\lambda_F,~\Omega=\omega/\omega_p$ and $\Omega_c=\omega_c/\omega_p$.
\begin{figure*}[ht]
\centering
\includegraphics[height=3in,width=7in]{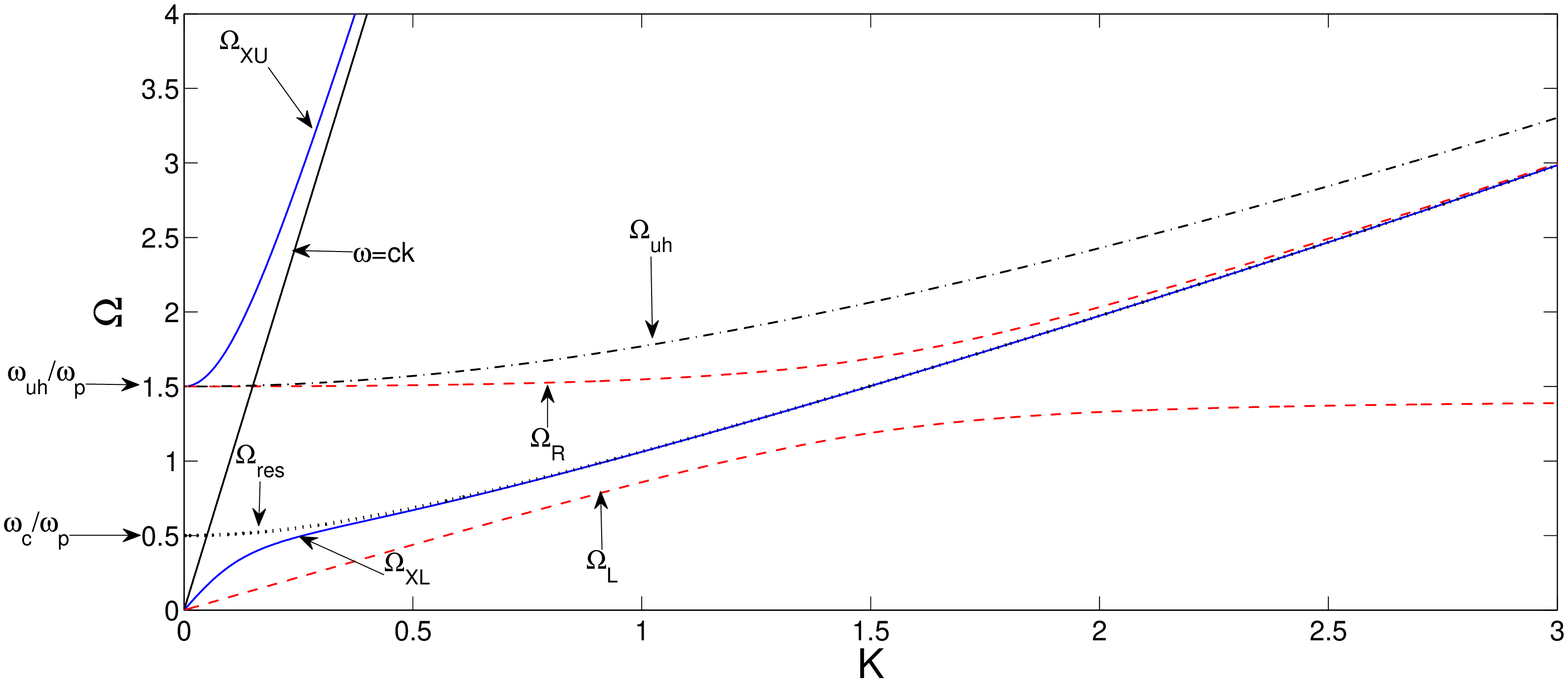}
\caption{The dispersion of the X-wave together with the cutoff and resonance frequencies are shown. While the solid lines (except the straight line which represents $\omega=ck$) represent the two branches of the X-wave ($\Omega_{XL}$ and $\Omega_{XU}$), the dashed lines correspond to the left- $(\Omega_L)$ and right-hand $(\Omega_R)$ cutoff frequencies. The dotted and dash-dotted lines, respectively, correspond to   the resonance frequency $(\Omega_{res})$ and the upper-hybrid wave frequency $(\Omega_{uh})$.    The parameter values are $\Omega_c=0.5$ and $H=0.2$. }
\label{fig1}
\end{figure*}
Note that in Eq. \eqref{dispersion-x-wave-dimless}, the influence of the quantum effects is mediated through the terms proportional to $\zeta_0$ both in the numerator and denominator. Here, we recall that the first term in $\zeta_0$  appears due to the degeneracy pressure of electrons and positrons, the second term is due to the exchange-correlation force and the third term is for the particle dispersion associated with the Bohm potential. Furthermore,  both the exchange correlation and the particle dispersive effects scale as $\sim H^2$. Thus, it  follows that in the long-wavelength EP plasma oscillations with $K\ll1$, the quantum dispersive effects can be negligible compared to the contribution from the exchange-correlation  force \cite{zamanian2015}. However,  for short-wavelength oscillations $(K\gg1)$, the effects can be reverse, i.e.,  the particle dispersion may be comparable to or  even dominate over the exchange-correlation force. However, for long-wavelength EP plasma oscillations in dense environments where $H<1$, the degeneracy pressure may play a dominating role   over the exchange-correlation force as well as the particle dispersion. In the next sections we will analyze these effects numerically on the wave dispersion as well on the phase and group velocities of the X  wave. 

Equation \eqref{dispersion-x-wave-dimless}, while expressed in $\Omega$, gives   two modes of the X-wave  
\begin{eqnarray}
\Omega_{XL,XU}^2=&&\frac{1}{2}\left[\Omega_X\right.\notag\\
&&\left.\mp\sqrt{\Omega_X^2-4\left( 2\zeta_0K^2+c_0^2K^2\Omega_{res}^2\right)}\right], \label{X1-X2}
\end{eqnarray}  
where $\Omega_X=\Omega_{res}^2+c_0^2K^2+2$ and $\Omega_{res}^2=\Omega_c^2+\zeta_0K^2$, to be shown as the square of the resonance frequency shortly.
\subsection{Phase velocity and group velocity of X-wave} \label{sec-sub-ph-gr}
The expressions for the phase velocity $(\omega/k)$ and the group velocity $(d\omega/d k)$ can be obtained from the dispersion relation  \eqref{dispersion-x-wave-dimless} by a straightforward algebra as
\begin{equation}
V_p\equiv \frac{\omega}{kV_F}=c_0\left(1-\frac{2}{\Omega^2}\frac{\Omega^2-\zeta_0 K^2}{\Omega^2-\Omega_{res}^2}\right)^{-1/2}, \label{phase-velo}
\end{equation}
\begin{equation}
V_g\equiv \frac{1}{V_F}\frac{d\omega}{dk}=\frac{\Omega_1\left(\zeta_0+H^2K^2/4\right)-\Omega_2c_0^2}{V_p\left(\Omega_1-\Omega_2\right)},\label{group-velo}
\end{equation}
where $\Omega_1=2-\Omega^2+c_0^2K^2$ and $\Omega_2=\Omega^2-\Omega_{res}^2$.
Note that corresponding to two different X-wave modes given by Eq. \eqref{X1-X2}, and using Eqs.  \eqref{phase-velo} and \eqref{group-velo} one can obtain different expressions for the phase velocity and group velocity.  In Sec. \ref{sec-results-discuss} we will see that while the group velocity of the branch $\Omega_{XL}$ increases with $K>1$, the upper branch $\Omega_{XU}$ approaches a constant value for $K>1$.
\subsection{Cutoff and resonance of X-wave} \label{sec-sub-cutoff-reso}
The dispersion relation of the X-wave is somewhat complicated. However, to analyze its properties it is useful to obtain the cutoff and resonance frequencies. Such cutoffs occur when the index of refraction $ck/\omega$ goes to zero or when the wavelength becomes infinite. 
On the other hand, the resonance of the X-wave occurs when the index of refraction becomes infinite or the wavelength becomes zero. The X-wave will then be reflected at the cutoff frequency and absorbed at the resonance. Thus, from Eq. \eqref{dispersion-x-wave-dimless}  the cutoff frequencies can be obtained as
\begin{equation}
\Omega_{R,L}^2=\frac{1}{2}\left[\Omega_{uh}^2\pm\sqrt{\Omega_{uh}^4-8\zeta_0K^2}\right], \label{cutoff-R-L}
\end{equation}
where $\Omega_{uh}^2=\Omega_c^2+\zeta_0K^2+2$ is the quantum modified hybrid frequency, and the suffices $R$ and $L$, corresponding to the plus and minus sign, respectively, stand for the right- and left-hand cutoff frequencies of the X-wave modified by the quantum effects. Note that in absence of the quantum effects, while the left-hand cutoff occurs at a zero value, the right-hand cutoff occurs at the upper-hybrid frequency.  The resonance frequency is obtained from Eq. \eqref{dispersion-x-wave-dimless} as
\begin{equation}
\Omega_{res}^2=\Omega_c^2+K^2\left(1-3.2H^2+\frac{1}{4}H^2K^2\right). \label{resonance}
\end{equation}
Thus, the resonance frequency of the X-EM wave is also modified by the quantum effects, in absence of which the resonance occurs at the cyclotron frequency instead of the upper-hybrid frequency as in classical X-Em waves \cite{goldston1995}. 
\section{Results and discussion}\label{sec-results-discuss}
We numerically investigate the dispersion properties of the X-wave [Eq. \eqref{X1-X2}] through the behaviors of the cutoff and resonance frequencies given by Eqs. \eqref{cutoff-R-L} and \eqref{resonance}. The latter are important in part as they define the pass and stop bands where the X-EM waves can propagate in EP plasmas. These are illustrated in Fig. \ref{fig1}. It clearly shows that the X waves (solid lines except that represents $\omega=ck$) can propagate in the pass band regions of $\Omega_L<\Omega<\Omega_{res}$ and $\Omega>\Omega_{uh}$, however, a stop band exists in the range $\Omega_{res}<\Omega<\Omega_R$. The cutoffs (dashed lines) are at the frequencies where the dispersion relation vanishes into the refractive index $ck/\omega=0$, and the resonances (dotted line) are where $ck/\omega\rightarrow\infty$. We also note that while for the upper pass band $V_p>c_0$,  the lower  branch of the X-wave propagates with a phase velocity $(V_p)$ smaller than the speed of light in vacuum $(c_0)$. Furthermore, because of the quantum effects, namely the degeneracy pressure, the exchange-correlation and the particle dispersion, the upper-hybrid frequency, and the cutoff and resonance frequencies are no longer  constants but are dispersive and increase with the wave number $K$. Furthermore, it is also seen from Fig. \ref{fig1} that in contrast to classical EM wave, the group velocities of the X waves do  not tend to vanish in the vicinity of the upper-hybrid frequency.  
\par
\begin{figure*}[ht]
\centering
\includegraphics[height=2.5in,width=7in]{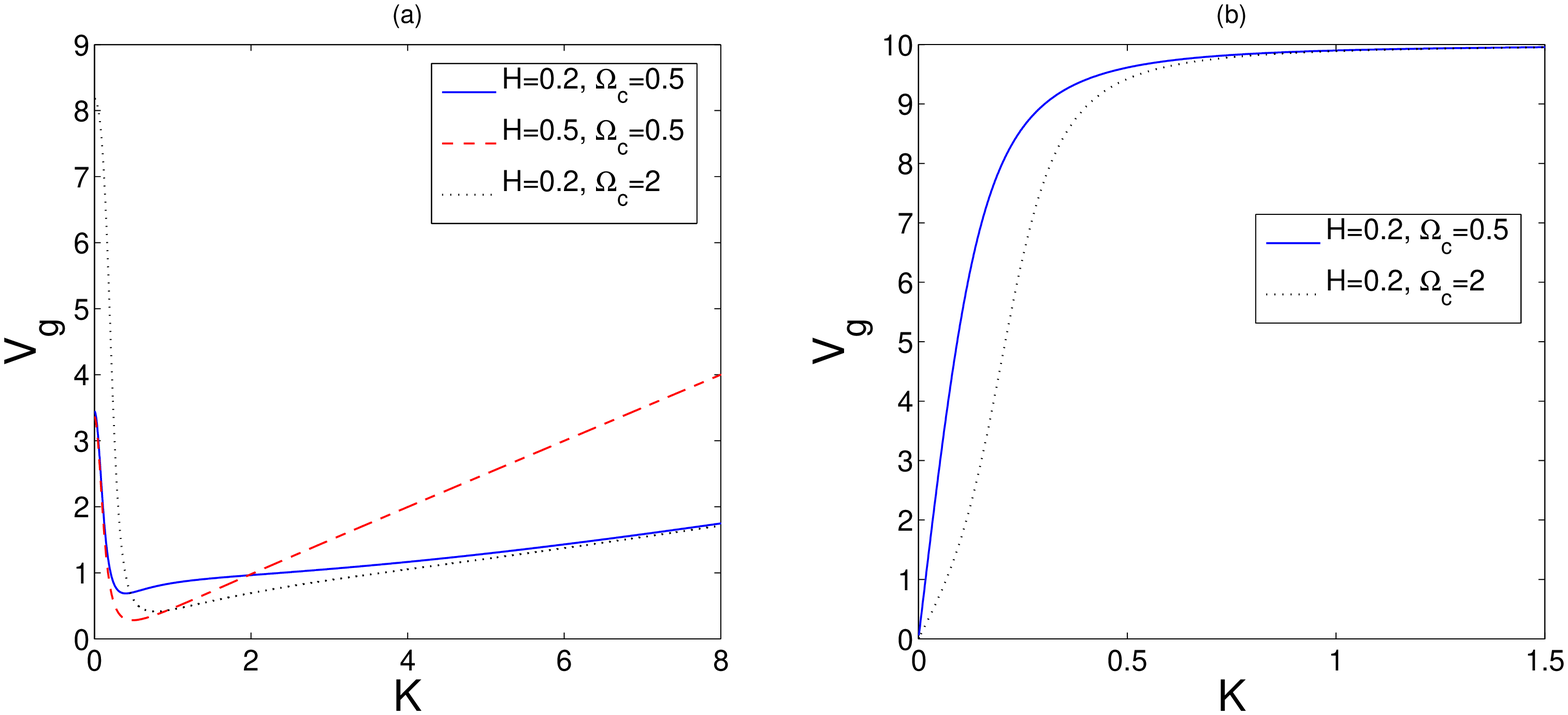}
\caption{Group velocities of the X wave [Eq. \eqref{group-velo}] are plotted against the wave number $K$ for different values of the quantum parameter $H$ and the cyclotron frequency $\Omega_c$ as in the legends. The left and right panels (a) and (b) are corresponding to  the  lower and upper branches   $\Omega_{XL}$ and $\Omega_{XU}$ of the X-wave respectively.}
\label{fig2}
\end{figure*}
The properties of  the group velocities of the  two branches of the X wave are shown in Fig. \ref{fig2} for different values of the quantum parameter $H$ and the cyclotron frequency $\Omega_c$. It is seen that while the parameter $H$ does not have any influence on the upper branch $\Omega_{XU}$ [see the right panel (b)], it can reduce the group velocity of the lower branch $\Omega_{XL}$ in the range $0<K<2$ and increase in the other regime [see the left panel (a)]. Also, $V_g$ of the branch $\Omega_{XU}$ approaches a constant value, whereas that of $\Omega_{XL}$ keeps increasing with   $K>1$. Furthermore, the effect of the cyclotron frequency on $V_g$ of  $\Omega_{XL}$ is to increase its value in the regime of $0<K\lesssim1$ and decrease in $K\gtrsim1$. However, increasing values of the same $(\Omega_c)$ decreases the values of   $V_g$ of  $\Omega_{XU}$.
\par
\begin{figure*}[ht]
\centering
\includegraphics[height=2.5in,width=7in]{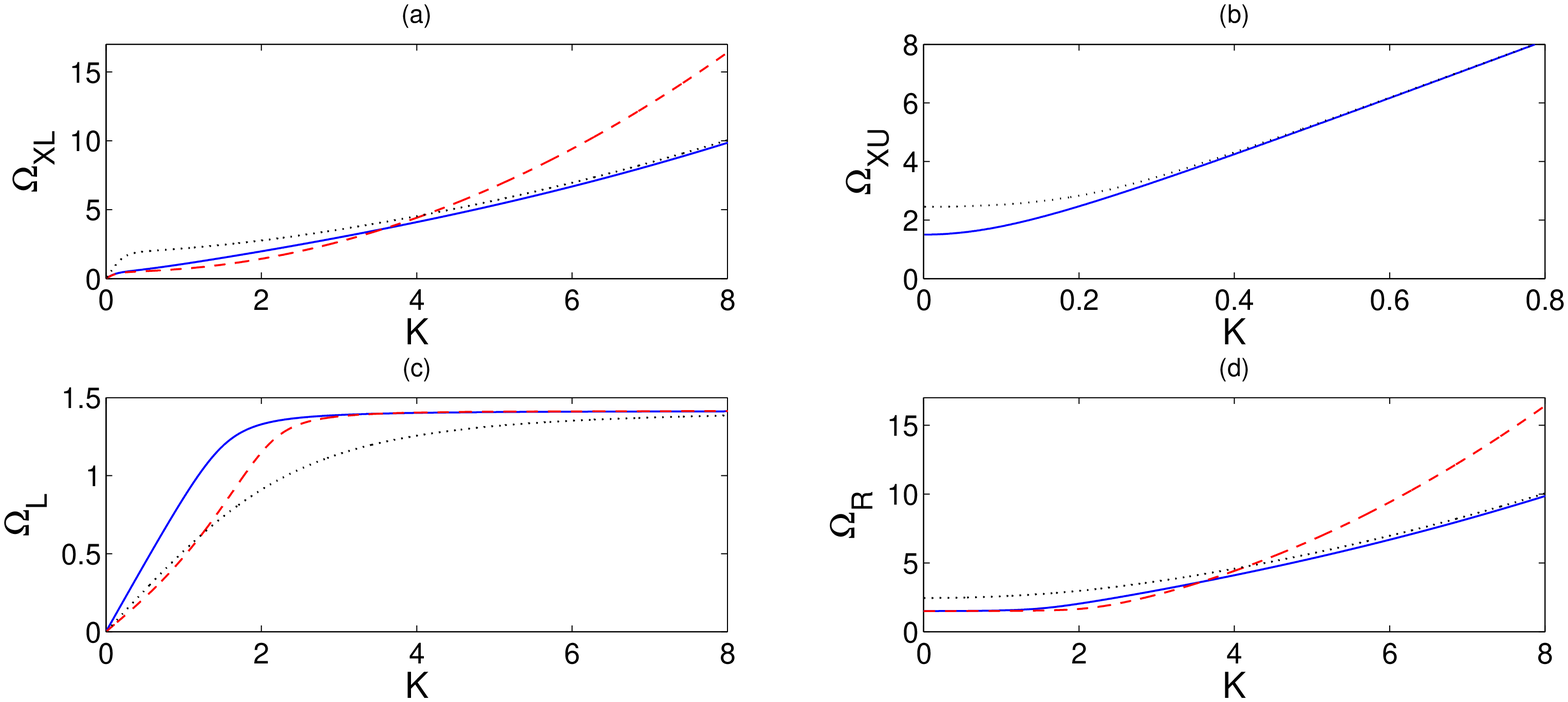}
\caption{Two branches  $(\Omega_{XL},~\Omega_{XU})$  of the X-wave [see the upper panels (a) and (b)] and two cutoff frequencies $(\Omega_{L},~\Omega_{R})$ [see the lower panels (c) and (d)]      are plotted  against the wave number $K$ for different values of the quantum parameter $H$ and the cyclotron frequency $\Omega_c$ as in Fig. \ref{fig2}, i.e., $H=0.2,~\Omega_c=0.5$ (solid line); $H=\Omega_c=0.5$ (dashed line line) and $H=0.2,~\Omega_c=2$ (dotted line). It is seen that the quantum parameter $H$ does not have any significant effect on the upper branch $(\Omega_{XU})$ of the X-wave.}
\label{fig3}
\end{figure*}

   
Figure \ref{fig3} clearly shows how the quantum parameter $H$ and the the cyclotron frequency $\Omega_c$ greatly influence    the characteristics of the lower/upper branches  as well as the cutoff frequencies of the X-EM wave. It is seen that the values of $\Omega_{XU}$ remain  unchanged with $H$, whereas the other frequencies get significantly modified by this parameter. Note that in most the cases, except for $\Omega_L$, the effect of $\Omega_c$ is to increase the frequencies.    It is also seen that $\Omega_L$ approaches a constant value at large $K$ and the effects of $\Omega_c$ on the frequencies become significant in the regime of $K\rightarrow0$. It is also found that the behaviors of the lower branch $\Omega_{XL}$ of the X-wave are almost similar to those of the right-hand cutoff frequency $\Omega_R$ [see panels (a) and (d)].
\par
\begin{figure*}[ht]
\centering
\includegraphics[height=2.5in,width=7in]{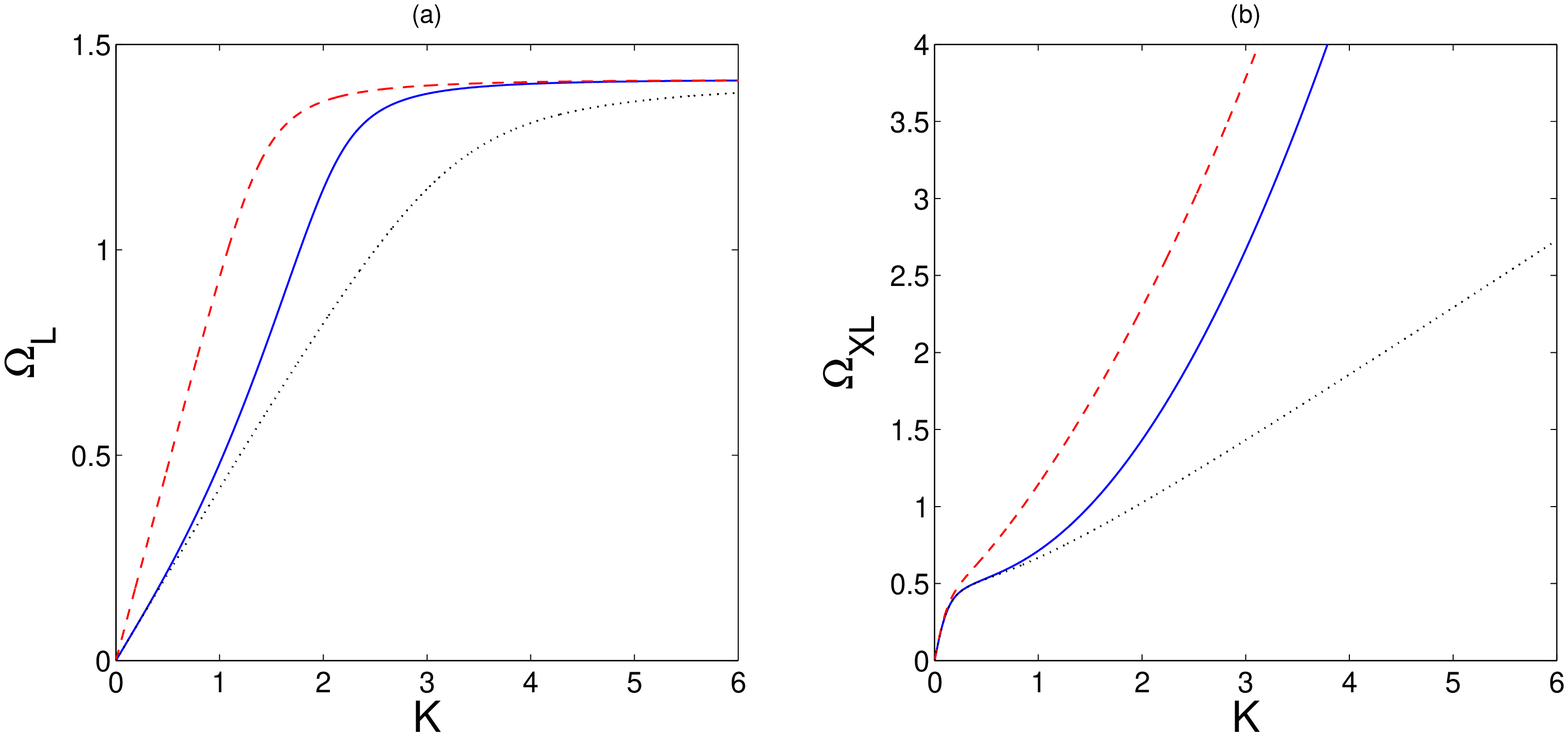}
\caption{The left-hand cutoff frequency [$\Omega_L$ in Eq. \eqref{cutoff-R-L}] and the lower branch of the X-wave [$\Omega_{XL}$ in Eq. \eqref{X1-X2}]  are plotted to show the effects of different quantum parameters. The solid line is such when all the three effects, namely the degenerate pressure, the exchange-correlation and the quantum dispersion are present ($H=\Omega_c=0.5$). The dashed line corresponds to the effects of the degenerate pressure and the quantum dispersion, and in absence of exchange-correlation effect. The dotted line corresponds to the effects of the degenerate pressure and the exchange-correlation force, and in absence of the quantum dispersion. Note that $\Omega_R$ has  similar behaviors as of $\Omega_{XL}$ with the quantum parameter $H$, and $\Omega_{XU}$ remains almost unchanged with $H$. }
\label{fig4}
\end{figure*}

To investigate the influence of different quantum forces separately and as  illustrations we have plotted $\Omega_L$ and $\Omega_{XL}$ (Here, we recall that $\Omega_{XL}$ has similar behaviors as of $\Omega_R$ and the quantum parameter $H$ does not have any significant effect on  $\Omega_{XU}$) with $K$ as in Fig. \ref{fig4}. It is shown that  for the left-hand cutoffs the correlation effect can be negligible for higher values of $K$ or for short-wavelength oscillations $(K\gg1)$, however, its effect  is no longer  negligible for  moderate values of $K$ or as  $K\rightarrow0$ (see the solid and dashed lines). For the right-hand cutoff or the lower branch of the X-wave, the effect of the correlation force is significant in the range $K>1$. In both the panels (a) and (b) it is also seen that the particle dispersion dominates over the correlation force and the degeneracy pressure in the range $K>1$.
\section{Conclusion} \label{sec-conclusion}
We have investigated the dispersion properties of the X-EM waves in a magnetized degenerate EP-pair plasma with the effects of weakly relativistic degeneracy pressure, the quantum force associated with the Bohm potential and that due to exchange and correlations of electrons and positrons. It is shown that the latter two effects, which scale as $H$ (the ratio of plasmon energy to the Fermi energy densities) along with the degeneracy pressure significantly modify the refractive index of the X-wave. It is also  seen that while   the particle dispersion becomes  important for short-wavelength oscillations, the exchange correlation can no longer be negligible (or can even dominate over the other quantum effects) in the long-wavelength perturbations. The upper-hybrid frequency together with the cutoff and resonance frequencies are also shown to be dispersive as modified by the quantum effects. Furthermore, in contrast to the classical X-EM waves, the group velocity is shown to be non-vanishing in the vicinity of the upper-hybrid frequency. To conclude, the results should be useful for understanding the dispersion properties of X-EM waves that can propagate in magnetized dense EP-pair plasmas such as those in magnetized white dwarf stars, neutron stars etc.

\acknowledgments{This work was supported by UGC-SAP (DRS, Phase III) with Sanction order No.
F.510/3/DRS-III/2015(SAPI), and UGC-MRP with F. No. 43-539/2014
(SR) and FD Diary No. 3668.}

\end{document}